\documentclass[sigconf]{acmart}

\acmConference[arXiv Preprint]{}{2025}{}
\acmBooktitle{}
\renewcommand\footnotetextcopyrightpermission[1]{} 

\usepackage{algorithmic}         
\usepackage{graphicx}            
\usepackage{textcomp}            
\usepackage{xcolor}              
\usepackage{tabularx}            
\usepackage{array}               
\usepackage{booktabs}            
\usepackage{multirow}            
\usepackage{float}               
\usepackage{stfloats}            
\usepackage{subcaption}          
\usepackage{fontawesome}         
\usepackage{enumitem}            
\setlist[itemize]{leftmargin=*}  

\usepackage{listings}
\usepackage{framed} 
\colorlet{acmboxrule}{gray!75!black}
\colorlet{acmboxbg}{gray!5}
\makeatletter
\newenvironment{acmbox}{%
  \MakeFramed{\advance\hsize-\width \FrameRestore}%
}{\endMakeFramed}
\makeatother

\begin{document}

\title{PromptDebt: A Comprehensive Study of Technical Debt Across LLM Projects}

\author{Ahmed Aljohani}
\email{ahmedaljohani@unt.edu}
\affiliation{
  \institution{University of North Texas, TX, USA}  
\city{}
\state{}
\country{}
}

\author{Hyunsook Do}
\email{hyunsook.do@unt.edu}
\affiliation{
  \institution{University of North Texas, TX, USA}  
\city{}
\state{}
\country{}
}






\begin{abstract}
Large Language Models (LLMs) are increasingly embedded in software via APIs like OpenAI, offering powerful AI features without heavy infrastructure. Yet these integrations bring their own form of self-admitted technical debt (SATD). In this paper, we present the first large-scale empirical study of LLM-specific SATD: its origins, prevalence, and mitigation strategies. By analyzing 93,142 Python files across major LLM APIs, we found that 54.49\% of SATD instances stem from OpenAI integrations and 12.35\% from LangChain use. Prompt design emerged as the primary source of LLM-specific SATD, with 6.61\% of debt related to prompt configuration and optimization issues, followed by hyperparameter tuning and LLM-framework integration. We further explored which prompt techniques attract the most debt, revealing that instruction-based prompts (38.60\%) and few-shot prompts (18.13\%) are particularly vulnerable due to their dependence on instruction clarity and example quality. Finally, we release a comprehensive SATD dataset to support reproducibility and offer practical guidance for managing technical debt in LLM-powered systems. 


\end{abstract}


\maketitle

\section{Introduction}
\label{sec:Introduction}

Large Language Models (LLMs) have rapidly become integral to modern software development~\cite{kaddour2023challenges}, enabling developers to perform a wide range of downstream tasks, including text summarization, conversational agents, and advanced automation~\cite{hou2024large, chen2024empirical}. To cope with the substantial cost and complexity of training and maintaining in-house LLMs, developers increasingly rely on APIs from companies such as OpenAI, Anthropic, and Cohere for seamless integration and rapid development~\cite{openai2024overview, anthropic2024gettingstarted, cohere2024apireference}.

The integration of LLMs into applications introduces new code-level artifacts \cite{tafreshipour2024prompting, hassan2024rethinking}, including prompts\cite{fan2023large}, hyperparameter settings \cite{arora2024optimizing} (e.g., temperature and max tokens), and LLMs frameworks like LangChain\cite{langchain2024apireference}. Among these, \textit{prompts} play a pivotal role as the primary mechanism for guiding LLM behavior, allowing developers to tailor outputs to specific needs through a process known as prompt engineering~\cite{sahoo2024systematic}.
In addition to these code-level artifacts, LLM-based applications encompass additional considerations, including operational costs, such as token usage\cite{openai2024tokenCost}, and advanced capabilities such as model fine-tuning offered by LLM APIs~\cite{openai2023fine-tuning}.

Although LLM APIs provide convenient access to powerful models, they pose significant long-term maintenance challenges. These challenges include the need for continuous adjustments to well-engineered prompts in response to model updates~\cite{ma2024my}, as well as the frequent modification of prompts—such as adding new instructions, rephrasing details, or restructuring templates—to align with evolving project requirements and the introduction of new features~\cite{tafreshipour2024prompting}. Furthermore, developers often resort to short-term fixes, such as prompt tweaks or token adjustments, to address the immediate challenges posed by LLMs rather than implementing scalable, long-term solutions~\cite{nahar2024beyond}. Over time, these continual adjustments risk accumulating \textit{Technical Debt (TD)}, increasing the maintenance burden required to keep LLM-based applications functional and aligned with user needs.

The technical debt (TD) concept describes the trade-offs developers make to accelerate software delivery at the expense of long-term maintainability~\cite{Cunningham1992}. Such trade-offs yield an \textit{interest} that must be paid in the form of future rework~\cite{potdar2014exploratory, wehaibi2016examining}. When these debts remain unaddressed, they become more expensive to resolve and can hinder the agility of the development team~\cite{wehaibi2016examining}. Within TD, \textit{Self-Admitted Technical Debt (SATD)} refers to cases in which developers explicitly acknowledge design or implementation deficiencies through the comments, often annotated with \texttt{TODO} or \texttt{FIXME}~\cite{potdar2014exploratory}. SATD offers valuable insights into technical debt's human-driven decisions, revealing where and why debt accumulates. Identifying SATD comments allows developers to prioritize areas in the codebase requiring immediate attention \cite{bavota2016large} and assist developers in controlling long-term maintenance costs\cite{maldonado2017empirical}.

Numerous studies have explored SATD in traditional software systems, identifying SATD types, developing automated detection tools, and analyzing their lifecycle~\cite{maldonado2017empirical, bavota2016large, wehaibi2016examining}. However, as software development continues to evolve, new application domains and emerging technologies introduce novel forms of SATD and associated challenges. In particular, Machine Learning (ML) and Deep Learning (DL) systems introduce unique complexities that go beyond those encountered in traditional software systems~\cite{sculley2015hidden}. For example, O'Brien et al.\cite{obrien202223} proposed a taxonomy of 23 ML-specific SATD types such as data dependency debt, model debt, and evaluation debt, while Bhatia et al.\cite{bhatia2023empirical} pinpointed highly debt-prone stages in ML pipelines like data preprocessing and model training. Moreover, Pepe et al.~\cite{pepe2024taxonomy} identified SATD categories specific to DL, highlighting infrastructure dependencies and inference-related SATD.

LLM-based applications introduce a fundamentally different development environment. Instead of managing large-scale training pipelines with dedicated hardware, developers leveraging LLM APIs primarily focus on \textit{prompt engineering}, \textit{hyperparameter tuning}, \textit{LLM framework integration}, and \textit{cost management}~\cite{chen2024empirical}. Yet we lack data on how these practices contribute to SATD in real-world LLM projects. Specifically, developers of LLM-based applications:

\vspace*{-3pt}
\begin{enumerate}
    \item Prioritize \textbf{prompt management}, which is a critical aspect that previous studies have largely overlooked but is integral to shaping model behavior, defining user roles, aligning task contexts, and ensuring appropriate output formats. For instance, developers must decide whether to \texttt{ \#break new prompts into configurations} or use frameworks like LangChain to streamline prompt management (e.g., \texttt{\#TODO: use langchain?}).
    \item Utilize \textbf{prompt learning}\cite{brown2020language} techniques (e.g., in-context learning, such as \textit{n-shot} examples) as a more flexible alternative to data-intensive fine-tuning. A common challenge here includes SATD comments like \texttt{\#TODO: consider few-shot examples}.
    \item Iteratively configure \textbf{LLM hyperparameters} (e.g., temperature, top-p, and max tokens), which are crucial for optimizing LLM outputs~\cite{ronanki2024prompt} yet these configuration decisions are rarely captured in SATD studies~\cite{pepe2024taxonomy}. For instance, developers document issues like \texttt{\#TODO handle top\_p, top\_k, etc.} as unresolved configuration decisions.
    \item Adopt \textbf{LLM-specific frameworks} to orchestrate interactions between applications and LLMs. These frameworks also introduce their own SATD when decisions on implementation or customization are left suboptimal or incomplete.
    \item Manage \textbf{operational costs}, including token usage and model pricing, which are unique to the LLM domain and add a distinct dimension to technical debt.
\end{enumerate}

\vspace*{-3pt}

By identifying these distinct sources of SATD in LLM-based applications, we aim to enhance existing ML/DL taxonomies~\cite{pepe2024taxonomy, obrien202223, bhatia2023empirical, liu2020using}, systematically identify poor prompt engineering practices such as ``prompt smells''~\cite{ronanki2024prompt} and ``prompt requirement smells''~\cite{vogelsang2025impact}, which can compromise model performance, and motivate the development of automated tools for effective prompt refactoring. Our findings contribute to establishing best practices for the long-term maintainability and reliability of LLM-based applications. To achieve these goals, we conduct an empirical study to address the following research questions:

\vspace*{3pt}
\noindent    
    \textbf{RQ1: What is the extent and distribution of SATD in LLM projects?}
    \textit{\textbf{Motivation:}} As LLMs become integral to advanced applications, understanding the prevalence and patterns of SATD is crucial for identifying risks, improving maintainability, and developing targeted strategies to manage and mitigate technical debt in LLM-based projects effectively.
    
\vspace*{3pt}
\noindent
    \textbf{RQ2: Which parts of LLM-based applications are prone to SATD?}
    \textit{\textbf{Motivation:}} 
    LLM systems introduce complexities distinct from traditional ML/DL pipelines. Understanding which components most frequently incur SATD can help refine and extend existing technical debt taxonomies.
    
\vspace*{3pt}
\noindent    
    \textbf{RQ3: Which prompt techniques are more prone to SATD?}
    \textit{\textbf{Motivation:}} Building on insights from RQ2, this question further investigates prompt-level debts to identify which specific prompt techniques\cite{pister2024promptset} might introduce unique technical debt challenges.\\

This paper advances the understanding of SATD in LLM-based applications by:
\begin{enumerate}

\item Conducting the first empirical study to identify and classify five unique types of LLM-related SATD, analyzing their prevalence and impact.

\item Providing actionable guidelines for effective mitigation strategies to reduce LLM-related SATD.    

\item Providing a comprehensive replication dataset to support reproducibility and facilitate further research.\footnote{\url{https://doi.org/10.5281/zenodo.15292881}}\\

\end{enumerate}

The paper is structured as follows: Section~\ref{sec:ResearchMethod} presents the approach used in this study, and Section \ref{sec:StudyResults} explains the results and findings of the study. Section \ref{sec:Study_Implication} presents the implications and additional discussion, and Section \ref{sec:ThreatsToValidity} discusses the limitations of our study. Section \ref{sec:RelatedWork} presents related work of self-admitted technical debt and their effect on traditional and ML software projects. Finally, in Section~\ref{sec:Conclusion}, we provide conclusions and discuss future work.

\section{Related Work} 
\label{sec:RelatedWork}

In this section, we discuss existing work relevant to self-admitted technical debt (SATD) research conducted on traditional and ML software projects.

\subsection{SATD in traditional software}
SATD has been widely studied in traditional software development. Potdar et al. \cite{potdar2014exploratory} found that 2.4\% to 31\% of codebases in open-source projects contain SATD, often introduced by experienced developers. Maldonado et al. \cite{maldonado2017empirical} showed that SATD is typically used to mark areas needing future improvements or bug fixes, remaining in the code for an average of 18 to 172 days, with design debt being the most common type \cite{maldonado2015detecting}. Further research by Bavota et al. \cite{bavota2016large} and Wehaibi et al. \cite{wehaibi2016examining} explored the relationship between SATD and software quality, showing that SATD tends to accumulate over time and has no direct correlation with code complexity.

\subsection{SATD in ML projects}

\textbf{SATD Prevalence and Categorization in ML Systems:} One of the key empirical studies on SATD in ML systems, Obrien et al.\cite{obrien202223} introduced a comprehensive taxonomy of 23 distinct types of SATD specifically tailored to ML systems. This work provides a deeper understanding of the types of debt that commonly arise in ML projects, including data preprocessing debt, algorithm-specific debt, and configuration debt. These types of debt are especially prevalent in the early stages of the ML pipeline, such as data preprocessing and model training, where developers must balance speed and accuracy. 

Building on this foundation, Bhatia et al. \cite{bhatia2023empirical} highlights the high prevalence of SATD in these projects, reporting that ML systems accumulate 2.1 times more SATD than non-ML software. This increased prevalence is largely attributed to the inherent complexities of data dependencies, configuration management, and the need for frequent experimentation in ML pipelines. As developers iterate quickly to refine models and manage large datasets, they often introduce SATD in areas like model performance, code maintainability, and data handling.

\textbf{SATD in Deep Learning Frameworks:} Another critical aspect of SATD in DL projects arises from the frameworks used to develop these models. Liu et al.\cite{liu2020using} investigated the prevalence of technical debt in deep learning frameworks such as TensorFlow and PyTorch. The study revealed that design debt and data dependency debt were particularly common in these frameworks, driven by the need for compatibility with evolving libraries and hardware infrastructure. The accumulation of SATD in DL projects often necessitates continuous refactoring to maintain code quality. Tang et al.\cite{tang2021empirical} explored the relationship between refactorings and SATD in ML/DL systems, demonstrating that refactoring efforts are frequently concentrated in areas where SATD is most prevalent, such as data preprocessing and model training.

Similar to Obrien et al.\cite{obrien202223}, Pepe et al.\cite{pepe2024taxonomy} introduced a comprehensive taxonomy of SATD specific to Deep Learning (DL) systems, categorizing 41 unique SATD types into infrastructure-related (e.g, such as GPUs and TPUs) and DL life-cycle-related debt. The DL life-cycle category focuses on SATD arising from data preparation, model design (e.g., attention masks and pooling layers), training logic (e.g., loss functions), and inference (e.g., post-processing). 
While previous efforts~\cite{pepe2024taxonomy} highlight SATD across the DL pipeline -- including prompts in inference-related SATD, their focus remains on DL system development rather than the unique challenges faced by developers building applications on top of LLM. 
In many DL systems, inference revolves around passing input data to a trained model, coupled with occasional adjustments to hyperparameters such as beam size or repetition penalties. By contrast, prompts in LLM-based applications serve as elaborate programming instructions\cite{reynolds2021prompt, liang2024prompts} that specify user roles, task contexts, and expected output formats. Refining these prompts often requires manual trial-and-error, multiple iterations, and can incur significant time and financial costs \cite{hassan2024rethinking}. This expanded function demands ongoing, fine-grained maintenance—for example, configuring n-shot examples or setting system/user prompts. Developers may defer these refinements (e.g., \texttt{\# TODO Figure out how to create system prompt}), and thus introducing technical debt that requires dedicated management. This distinction positions our work to explore SATD in LLM-based applications that remain unexplored in the existing literature.

\section{Research Method}
\label{sec:ResearchMethod}

\begin{figure*}[h!]
    \centering
    \includegraphics[width=1\textwidth]{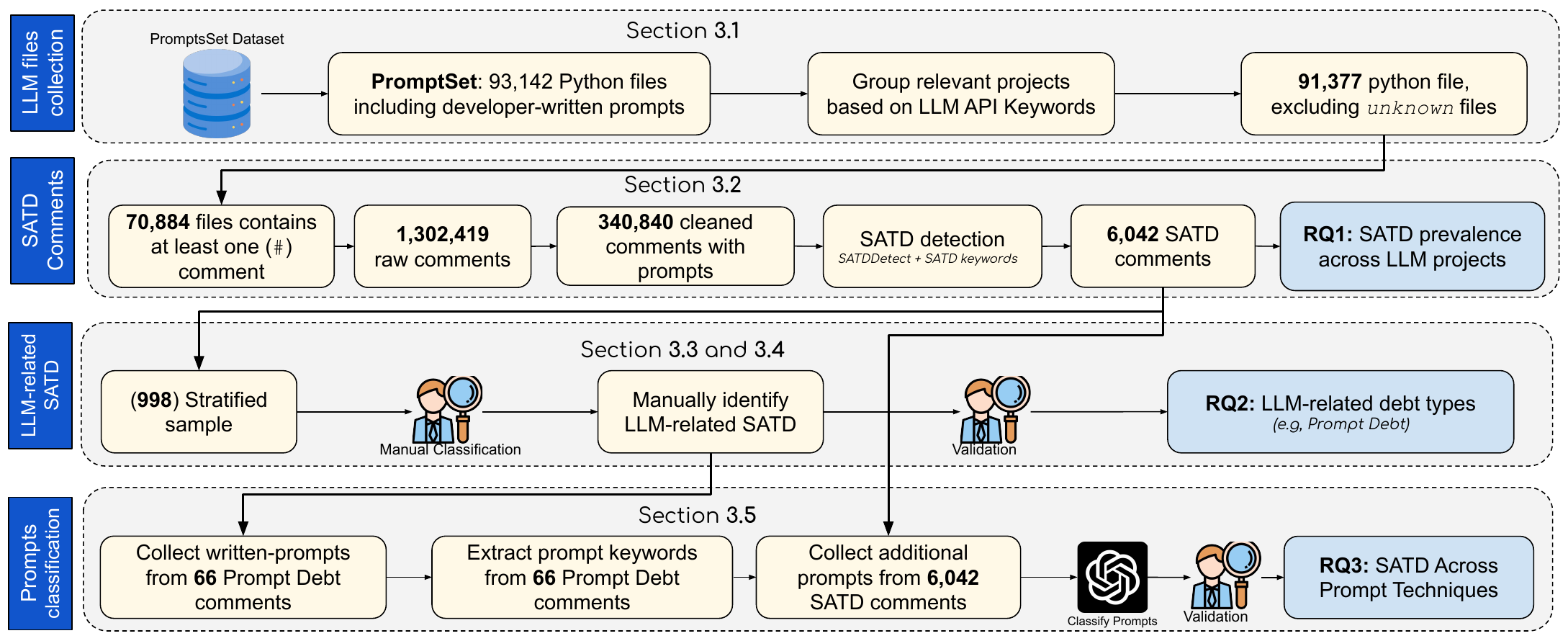} 
    \caption{An overview approach.}
    \label{fig:ApporachOverview}
\end{figure*}

Figure ~\ref{fig:ApporachOverview} illustrates the overall methodology used in our study. The approach involves multiple key steps: first, the collection of relevant Python files related to Large Language Models (LLMs) (the top layer in the figure); the second step involves extracting source code comments, conducting a thorough cleaning process, and detecting Self-Admitted Technical Debt (SATD) comments as shown in the second layer. In the third layer, we illustrate our manual classification of LLM-related SATD using our sample. Finally, we present the collection of developer-written prompts and their corresponding SATD instances.

\begin{table}[ht]
\centering
\begin{tabular}{ll}
\hline
\textbf{LLM} & \textbf{Count (\%)} \\
\hline
OpenAI & 41,403 (44.45\%) \\
LangChain & 22,348 (24.00\%) \\
LangChain, OpenAI & 20,803 (22.33\%) \\
Cohere & 3,517 (3.78\%) \\
Unknown & 1,765 (1.89\%) \\
Anthropic & 505 (0.54\%) \\
Cohere, OpenAI & 531 (0.57\%) \\
Anthropic, OpenAI & 446 (0.48\%) \\
Cohere, LangChain, OpenAI & 443 (0.48\%) \\
Anthropic, LangChain & 415 (0.45\%) \\
Cohere, LangChain & 331 (0.36\%) \\
Anthropic, LangChain, OpenAI & 307 (0.33\%) \\
Anthropic, Cohere, OpenAI & 154 (0.17\%) \\
Anthropic, Cohere, LangChain, OpenAI & 140 (0.15\%) \\
Anthropic, Cohere, LangChain & 24 (0.03\%) \\
Anthropic, Cohere & 10 (0.01\%) \\
\hline
\textbf{Total} & \textbf{93,142 (100.00\%)} \\
\hline
\end{tabular}
\vspace*{4pt}
\caption{Distribution of Files by LLM/Framework}
\vspace*{-10pt}
\label{tab:api_configuration}
\end{table}

\vspace*{-3pt}
\subsection{LLM Files Collection}
We utilized the PromptSet dataset \footnote{\url{https://huggingface.co/datasets/pisterlabs/promptset}}~\cite{pister2024promptset}. This dataset is curated to capture developer-written prompts involving major LLM libraries, such as \texttt{OpenAI}, \texttt{Anthropic}, and \texttt{Cohere}. It also includes references to frameworks like \texttt{LangChain}, which, while not an LLM itself, developers typically interact with the LLMs via such a framework rather than calling the LLMs APIs directly\cite{langchain2023}. By including LangChain in our SATD study, we aim to capture a more comprehensive view of how technical debt arises from the LLMs and the frameworks that manage LLM's input and output. 

The dataset, compiled in January 2024, was sourced by mining 37,944 GitHub repositories and focused on Python files that explicitly reference these LLM APIs or frameworks. It consists of 93,142 Python files, each containing source code and metadata, including developer-written prompts. To better understand the distribution of SATD within the dataset, we grouped the files according to the specific LLM libraries or frameworks they referenced (e.g., OpenAI, Anthropic, Cohere, and LangChain). This classification was based on unique keywords outlined by Pister et al.\cite{pister2024promptset}. In Table \ref{tab:api_configuration}, we present the LLM distribution. The complete list of keywords used for this grouping is provided in Appendix A.

Files that did not match the predefined API or framework keywords were classified under the \textit{Unknown} category. A manual examination of 20 sample files from this category revealed that they typically served as utility files that complemented LLM-related code, rather than directly interacting with the specified LLMs or frameworks. Consequently, we excluded these utility files from our analysis, resulting in a refined dataset of \textbf{91,377} Python files.

\vspace*{-3pt}
\subsection{SATD Comments}
In this section, we detail the process of extracting Python comments from LLM-related files and outline the steps taken to clean these comments for effective SATD detection. Additionally, we explain how SATDDetector, complemented by SATD-specific keywords, was utilized to identify SATD comments within the dataset.

\textbf{Comments Extraction:} To begin quantifying SATD comments, we first extracted source code comments from the 91,377 Python files in our dataset. This extraction was performed using Python’s \texttt{tokenize} module\footnote{\url{https://docs.python.org/3/library/tokenize.html}}, which breaks down Python source code into individual components (tokens), including single-line comments (\texttt{\#}). During the extraction process, we discovered that 20,493 out of the 91,377 Python files contained no single-line comments. As a result, these files were excluded from further analysis, reducing our dataset to 70,884 Python files. The initial comments extraction yielded 1,302,419 raw comments across the remaining files. To ensure the comments were suitable for SATD detection, we applied a cleaning process similar to those described in prior research \cite{obrien202223, da2017using}. This process involved removing non-informative or irrelevant comments, such as duplicate comments, license headers, generated code comments, and non-English comments. After cleaning, we obtained a refined dataset of 340,840 comments, which became the basis for SATD analysis.

\textbf{Detecting SATD:} For SATD detection, we adopted a methodology by O'Brien et al.\cite{obrien202223}. We first employed the SATDDetector tool\footnote{\url{https://github.com/Tbabm/SATDDetector-Core}}, an automated tool designed to classify SATD comments based on a machine learning model trained on traditional software project data \cite{huang2018identifying}. SATDDetector, however, was primarily trained on comments from traditional software projects and showed some false positives when applied to ML-related SATD, as noted by O'Brien et al. To mitigate this issue, we supplemented SATDDetector's results with an additional keyword-based detection approach, using SATD-related keywords identified in  \cite{da2017using}. From the 340,840 cleaned comments, we detected a total of \textbf{6,042} SATD comments, representing approximately 1.77\% of the entire comment dataset. Further analysis showed that 3,630 (5.12\%) of the 70,884 Python files contained at least one SATD comment.

\vspace*{-3pt}
\subsection{SATD Sampling}
To address our RQ2, we utilized a statistically significant, stratified sample of 998 SATD comments (with a 95\% confidence level and a 5\% margin of error) from the 6,042 SATD comments identified in LLM-related files. Our sampling method followed the statistical approach described by \cite{bhatia2023empirical, obrien202223, zhang2019empirical}, where we ensured a proportional representation of comments across various LLM APIs. The sample distribution is detailed in Table \ref{tab:sample_sizes}.

\begin{table}[ht]
\centering
\begin{tabular}{lc}
\toprule
\textbf{LLM} & \textbf{Sample Size} \\
\midrule
OpenAI & 382 \\
OpenAI, LangChain & 250 \\
LangChain & 167 \\
Cohere & 85 \\
OpenAI, Cohere & 25 \\
OpenAI, Anthropic, LangChain & 21 \\
OpenAI, Anthropic & 15 \\
OpenAI, Cohere, LangChain & 15 \\
Anthropic & 10 \\
OpenAI, Anthropic, Cohere & 9 \\
Anthropic, LangChain & 8 \\
Cohere, LangChain & 4 \\
OpenAI, Anthropic, Cohere, LangChain & 3 \\
Anthropic, Cohere & 1 \\
\bottomrule
\end{tabular}
\vspace*{4pt}
\caption{Sample Sizes by LLMs}
\vspace*{-20pt}
\label{tab:sample_sizes}
\end{table}

\subsection{Manual Categorization of LLM-Related SATD}
We aim to extend existing ML/SATD taxonomies by identifying categories that are unique to LLM-based projects. To this end, we conducted a manual analysis of 998 comments from our dataset to construct a comprehensive classification of LLM-related SATD. Grounding our approach in \textbf{(i)} official documentation for widely used LLMs~\cite{openai2023, langchain2023, cohere2023, anthropic2023} and \textbf{(ii)} LLM application challenges and evolution~\cite{chen2024empirical, hou2024voices, tafreshipour2024prompting,pepe2024taxonomy}, we focused on practical aspects that frequently introduce debt in LLM-based systems—namely, prompting, fine-tuning, hyperparameter tuning, token cost, and framework usage (e.g., LangChain).

For instance, official LLM documentation~\cite{openai2023, langchain2023, cohere2023, anthropic2023} outlines recommended practices for prompt design, hyperparameter settings, and LLM usage, yet our dataset contained numerous developer comments (e.g., \texttt{\#TODO: figure out max input size for cohere}, \texttt{\#TODO: consider better prompt}) demonstrating a deviation from these guidelines. In addition, research findings~\cite{chen2024empirical, hou2024voices, tafreshipour2024prompting, pepe2024taxonomy}  highlight common pitfalls such as repeated prompt revisions and token-limitation workarounds. Observing similar challenges in our dataset allowed us to more precisely label and categorize instances of SATD related to LLM usage. Interestingly, while these sources did not explicitly mention LLM framework usage (e.g., LlamaIndex\cite{llamaindex2024overview}), our classification captured multiple SATD instances related to such frameworks.
From these analyses, we identified new categories to reflect these LLM-specific concerns. In Table~\ref{tab:llm_satd_types} (the table will be explained in detail in Section~\ref{sec:StudyResults}.2), we show the types we found based on our sample, their definition, and examples. 

To address the risk of bias in our manual classification of SATD comments, we adopted a multi-rater approach inspired by prior SATD studies \cite{obrien202223}. Two qualified investigators independently labeled each comment, and we computed Cohen’s Kappa\cite{mchugh2012interrater} to quantify inter-rater reliability. We obtained a Cohen’s Kappa of 0.74, indicating a substantial level of agreement. Any disagreements were resolved through a joint discussion with a third investigator.

\begin{table*}[htbp]
\small
\centering
\caption{LLM-Related SATD per comment, Definition and Examples}
\vspace*{-5pt}
\begin{tabular}{p{2.2cm} p{4.5cm} p{7.3cm} p{2cm}}
\toprule
\textbf{SATD Type} & \textbf{Definition} & \textbf{Examples} & \textbf{Count (\%)}\\
\midrule
Prompt Debt 
& Debt resulting from poorly structured, inefficient, or suboptimal prompts. 
& 
    \# TODO: refine the prompt template\newline
    \# TODO Figure out how to create system prompt\newline
    \# HACK since we can't specify exact prompt in .yaml\newline
    \# TODO Fix cases where this makes the prompt too long
& 66 (6.61\%) \\\hline

LLM-hyperparameters Debt 
& Debt related to improper or suboptimal tuning of LLM hyperparameters. 
& 
    \# TODO handle top\_p, top\_k, etc.\newline
    \# TODO: Make max\_tokens configurable \newline
    \# TODO: figure out which temperature is best for evaluation
& 45 (4.51\%) \\\hline

LLM-Integrated Framework Debt 
& Debt arising from the integration and orchestration of LLM frameworks. 
& 
    \# TODO: remove hardcoded context size once ... via langchain\newline
    \# TODO: this check has been switched to "false" after LlamaIndex deprecated LLMPredictors\newline
    \# TODO: Figure out pipeline prompts to avoid this
& 43 (4.31\%) \\\hline

Cost Debt 
& Debt refers to explicit issues related to token consumption and model pricing selection.
& 
    \# TODO calculate cost based on the whole prompt\newline
    \# TODO Figure out how to use 'gpt-3.5-turbo' instead, since it's 1/10 the price of davinci\newline
    \# TODO Assume a cost of \$0.00006 per token.
& 21 (2.10\%) \\\hline

Learning Debt* 
& Debt related to insufficient fine-tuning or prompt learning. 
& 
    \# TODO: push a finetuned gpt2 image to Huggingface and ..\newline
    \# TODO: it might be profitable to fine-tune Result statements to emphasize key skills\newline
    \# TODO: consider few-shot examples
& 6 (0.60\%) \\
\bottomrule
\end{tabular}
\label{tab:llm_satd_types}
\vspace*{-8pt}
\end{table*}

\begin{table*}[t!]
\centering
\small
\caption*{*Refers to both data-intensive fine-tuning (continuing training on a task-specific dataset) and prompt learning\cite{brown2020language}, where temporary learning methods (e.g., In-context learning) used to guide the model's output.}
\vspace*{-15pt}
\end{table*}

\vspace*{-5pt}
\subsection{Prompt Classification}
To determine which prompt techniques used by developers are most subjected to technical debt (RQ3), we categorized the prompts (e.g., Zero/Few shots, Chain-of-Thoughts (CoT)) according to Pister et al. \cite{pister2024promptset} prompt techniques classifications.

The findings from RQ2, which focused on LLM-related debts, revealed that Prompt Debt was the most prevalent type of debt in our sample. Specifically, out of 998 instances, 66 were linked to issues involving prompts. To build on this insight, we began by collecting developer-written prompts associated with these 66 cases of Prompt Debt. Then, we expanded our prompt sample size by developing search strings based on the prompt keywords identified from these 66 instances to gain a deeper understanding. The keywords used were \texttt{prompt, conversation, message, and query}. We applied these keywords across the entire dataset of 6,042 SATDs. This approach enabled us to identify an additional 383 SATD comments that are associated with the prompts crafted by developers, bringing the total to 449 SATD comments related to prompting, along with their developer-written prompts.

However, in our analysis of the prompt content, we observed two distinct methods of prompt presentation. Some files include the full prompt (e.g., \faCommenting{} \textit{Prompt: [``You are a bot designed to summarize a ....'']}), which aligns with our objective to classify the precise techniques developers employ. In contrast, other prompts adopt a placeholder-based approach, replacing the actual text with placeholders such as \texttt{['PLACEHOLDER': 'PLACEHOLDER']}. Since our core study centers on analyzing SATD comments, we focused on only those prompts that explicitly displayed their full content, leading us to exclude 294 placeholder-based prompts and retain a final set of 155 prompts for classification.

To classify these 155 prompts, we employed LLM-based approach similar to previous studies\cite{pister2024promptset, tafreshipour2024prompting}. Our prompt contains the definitions of each prompt technique alongside illustrative examples in a few-shot setting, allowing us to effectively handle longer prompts (more than 40 words) that include multiple techniques. We manually validated a random sample of 40 instances ($\approx 26\%$ of the dataset) to assess the reliability of the prompt technique classifications. Among these, 92.5\% (37/40) were correctly classified, while three cases (7.5\%) were misclassified. The misclassifications primarily involved distinguishing between Instruction-based prompting and Chain-of-Thought (CoT) prompting, where all misclassified cases involved instructional prompts being incorrectly labeled as CoT due to the presence of multiple sub-instructions.

Each prompt in our dataset maintains a one-to-many relationship with its associated SATD comments. Nonetheless, it is important to note that while these SATD comments reference \textit{prompting activities}—as identified by prompt-related keywords—they do not necessarily tie back to the specific prompt techniques developers used. We provide an example of this case in the RQ3 results.
 
\vspace*{-3pt}
\section{STUDY RESULTS}
\label{sec:StudyResults}
\vspace*{-2pt}
The study results present the prevalence of SATD comments across all LLM files, as shown in Table \ref{tab:satd_counts_api_combination}. Out of 6,042 SATD comments, we selected 998 representative sample comments proportionally based on the distribution of LLMs in our dataset as described in Section \ref{sec:ResearchMethod}-A. From these comments, we identified SATD comments related to LLMs (see Table~\ref{tab:llm_satd_types}). Each LLM-specific SATD type in the table is supported by at least one comment in our sample. We used a subset of 155 developer prompts to analyze prompt techniques, each accompanied by its associated SATD comment. We then classified these prompts based on the prompt technique. Each technique shown in Table \ref{tab:prompt_types} is represented by at least one written-prompt in our sample.

\subsection{\textbf{RQ1: What is the prevalence of SATD in LLMs files?}}
\vspace*{-2pt}
The prevalence of SATD across LLM APIs offers valuable insight into the scale and significance of these LLMs' challenges, helping to determine whether they deserve greater attention. 
Table \ref{tab:satd_counts_api_combination} illustrates the distribution of SATD among various LLM APIs, with OpenAI alone accounting for a substantial 54.49\% of all identified SATD. 
Following OpenAI, the combination of OpenAI with LangChain constitutes 20.39\% of the SATD. 

A closer examination of our sample—250 SATD comments specifically related to both OpenAI and LangChain—provides insights into each API's contribution to the overall technical debt. Within this subset, we noticed that LangChain contributes slightly more to SATD than OpenAI. This is also evident, where LangChain alone accounts for 12.35\% of the SATD, which adds a layer of complexity that developers need to address.

\begin{table}[ht]
\vspace*{5pt}
\centering
\small
\setlength{\tabcolsep}{6pt} 
\begin{tabular}{lcc}
\toprule
\textbf{LLM} & \textbf{SATD Occur.} & \textbf{Total \# of} \\
& & \textbf{Comments} \\
\midrule
OpenAI                        & 3292 (54.49\%)         & 197,380            \\
OpenAI, LangChain             & 1232 (20.39\%)         & 72,044             \\
LangChain                     & 746 (12.35\%)          & 30,751             \\
Cohere                        & 351 (5.81\%)           & 23,605             \\
OpenAI, Cohere                & 98 (1.62\%)            & 4,766              \\
OpenAI, Anthropic, LangChain  & 81 (1.34\%)            & 1,819              \\
OpenAI, Anthropic             & 58 (0.96\%)            & 2,400              \\
OpenAI, Cohere, LangChain     & 58 (0.96\%)            & 2,909              \\
Anthropic                     & 37 (0.61\%)            & 1,271              \\
OpenAI, Anthropic, Cohere     & 34 (0.56\%)            & 761                \\
Anthropic, LangChain          & 30 (0.50\%)            & 1,300              \\
Cohere, LangChain             & 12 (0.20\%)            & 995                \\
OpenAI, Anthropic, Cohere, LangChain & 11 (0.18\%)      & 756                \\
Anthropic, Cohere             & 2 (0.03\%)             & 42                 \\
\midrule
\textbf{Total}                & \textbf{6042 (100\%)}  & \textbf{340,840}   \\
\bottomrule
\end{tabular}
\vspace*{2pt}
\caption{SATD Occurrence by LLM.}
\label{tab:satd_counts_api_combination}
\vspace*{-10pt}
\end{table}

On the other hand, Cohere and Anthropic, whether used independently or in combination with different APIs, exhibit lower incidences of SATD, with Cohere contributing 5.81\% and Anthropic only 0.61\%. This difference may stem from their less frequent usage compared to OpenAI and LangChain, as evidenced by the substantially lower total comment volumes of 23,605 for Cohere and 1,271 for Anthropic, which results in fewer instances of SATD.

Overall, the high occurrence of SATD in OpenAI-based systems and LangChain highlights the need for proactive technical debt management, particularly given the widespread adoption of these tools.


\begin{acmbox}
\noindent 
\textbf{Summary of RQ1:} OpenAI exhibits the highest occurrence of SATD compared to other LLMs. Additionally, the frequent integration of the LangChain framework contributed to a higher SATD within LLM-related files. 
\end{acmbox}

\subsection{\textbf{RQ2: Which parts of the LLM are prone to SATD?}}
In RQ2, we categorize and analyze LLM-related technical debts to identify which parts of LLM, such as prompting and fine-tuning, are most prone to SATD. This section includes examples illustrating these debts and providing insights into common challenges developers face in LLM-based projects.

Table~\ref{tab:llm_satd_types} presents the distribution of LLM-related SATD per comment in our sample. \textit{Prompt Debt} is the most common, with 66 instances (6.55\%) out of 998 comments, indicating frequent challenges in designing and optimizing prompts. \textit{LLM-
hyperparameters
Debt} follows with 45 instances (4.46\%), highlighting issues related to crucial LLM tuning parameters like \texttt{temperature}. \textit{LLM-Integrated Framework Debt}, with 43 instances (4.27\%), points to difficulties in integrating and managing frameworks such as LangChain for LLM workflows. Lastly, \textit{Cost Debt} (21 instances, 2.10\%) and \textit{Learning Debt} (6 instances, 0.60\%) appear less frequently, showing that concerns about tokens costs and incomplete fine-tuning are relatively rare but still notable.

\textbf{Prompt Debt:}
In our analysis of \emph{Prompt Debt} comments, the most frequent issue involves incomplete \textit{prompt configurations}—that is, the need for prompt structures or templates to be dynamically adjustable in order to accommodate changing requirements and contexts (as indicated by examples in Table~\ref{tab:llm_satd_types}). Developers frequently struggle to create flexible, reusable prompts that can be universally applied without extensive hardcoding. For instance, the comment \faCommenting{} \textcolor{gray}{ \texttt{\#TODO: other conversation template}} highlights the absence of a centralized, standardized approach to conversation-based task, while \textcolor{gray}{ \texttt{\#TODO: dynamically set the city here in the prompt}} shows that certain context-specific (i.e, city) details remain manually embedded. 

A second significant challenge developers encounter revolves around \textit{optimizing prompt design}. Our dataset reveals that LLM developers often struggle to craft prompts that achieve desired outcomes, follow the correct format and manage prompt length, which can directly influence the LLM’s performance for a given task~\cite{marvin2023prompt}. For example, a comment such as \faCommenting{} \textcolor{gray}{ \texttt{\#TODO write custom prompt and parse it to get best results}} underscores the need for more precise, task-specific prompts. Similarly, comments referencing overly long prompts (e.g.,\texttt{.. prompt too long}) point to the difficulty of maintaining clarity and focus within the token limits of the model, potentially degrading comprehension and output accuracy~\cite{liu2024lost}. Another design challenge is getting the output in the right format. Developers sometimes desire the model to respond in JSON or another structured layout, but the prompt does not clearly state it. This leads to comments like \faCommenting{} \textcolor{gray}{\texttt{\#TODO: Turn response to JSON}}, which shows that formatting issues are often postponed.

\textbf{LLM-hyperparameter Tuning Debt:} In LLM-based systems, tuning models and adjusting their hyperparameters is critical to achieving optimal results \cite{arora2024optimizing, ouyang2024empirical}. However, improperly configured parameters, such as \texttt{temperature}, and \texttt{top\_p}, can lead to suboptimal results and technical debt. Table~\ref{tab:llm_satd_types} shows several examples of developers facing difficulties tuning parameters to fit their specific use cases. For instance, the comment \faCommenting{} \textcolor{gray}{ \texttt{\#TODO: figure out which temperature is best for evaluation }} demonstrates a common challenge where developers struggle to determine the appropriate temperature setting to control the randomness of the model's output. Similarly, \faCommenting{} \textcolor{gray}{ \texttt{\# TODO: handle top\_p, top\_k, etc.}} reflects the complexity in adjusting probability and sampling techniques, which can significantly impact the diversity and creativity of generated LLM responses \cite{arora2024optimizing}.

\textbf{LLM-Integrated Framework Debt}: frameworks like LangChain \cite{langchain2023} and LlamaIndex\cite{llamaindex2024overview} provide significant advantages for developers working with LLM-based applications. In our analysis, LangChain was the most frequent framework regarding associated technical debt, having the majority of SATD instances compared to other frameworks like LlamaIndex.

LangChain helps developers in two main perspectives: \textit{prompt management} and \textit{LLM output handling}. In prompt management, LangChain allows developers to create, format, and optimize prompts before sending them to an LLM. In LLM output handling, LangChain processes the responses from the LLM such as parsing and formatting the LLM output.

The primary challenges in using LangChain often come from \textit{prompt management} as shown in Table~\ref{tab:llm_satd_types}. For instance, \faCommenting{} \textcolor{gray}{ \texttt{\# TODO: Replace with f strings and all PromptTemplate}} indicates a situation to refactor prompt construction using Python's f-strings and LangChain's \texttt{PromptTemplate} \footnote{LangChain PromptTemplate Documentation: \url{https://python.langchain.com/api_reference/core/prompts/langchain_core.prompts.prompt.PromptTemplate.html}} for more precise, more maintainable prompt.
For LLM output formatting, technical debt is evident in comments like \faCommenting{} \textcolor{gray}{ \texttt{\# TODO: use langchaing output parser}}. This shows a recognized but postponed effort to incorporate LangChain’s output parser, which can lead to inefficient handling of responses, causing potential inconsistencies in downstream applications that rely on formatted and predictable outputs.

\vspace*{3pt}
\textbf{Cost Debt:} While LLMs are highly accessible and powerful, they come with significant cost considerations. Costs are broken down into input and output costs—both scaling with token length—making extensive use prohibitively expensive for LLM-based applications \cite{wang2024survey}.
We found that Cost Debt is strongly related to finding options for i) lower-cost models or ii) minimizing token usage. The examples in Table~\ref{tab:llm_satd_types} indicate that the developers are strongly aware of LLM usage costs. For instance, \faCommenting{} \textcolor{gray}{ \texttt{\# TODO Figure out how to use 'gpt-3.5-turbo' instead, since it's 1/10 the price of davinci}} illustrates the consideration of cost-effective model alternatives. Comments like \textcolor{gray}{ \texttt{\# TODO: Truncate the output to meet the token requirement and save\textdollar\textdollar}} highlight developers' efforts to minimize token usage. This indicates that cost management is a significant aspect of SATD in LLM projects, as developers actively look for ways to reduce expenses by minimizing token usage or finding lower-cost models. 

\textbf{Learning Debt:} Developers use prompts to guide LLMs in generating desired outputs, often employing learning techniques such as in-context learning, including few-shot prompting\cite{brown2020language}. Another effective approach is fine-tuning LLMs on task-specific data. Currently, certain models by OpenAI can be fine-tuned to better align with specific datasets \cite{openai2023fine-tuning}. Fine-tuning typically allows developers to achieve more consistent and high-quality results compared to using prompts alone. We consider two learning approaches. The first approach is fine-tuning, which involves training a pre-trained model on a new dataset tailored to a desired task \cite{radford2019language}. The second approach is prompt-learning \cite{brown2020language}. In this method, the model is provided with n-examples at inference time, guiding it to perform the task based on the context provided by these examples. 

Our analysis of Learning Debt indicates that insufficient or incomplete fine-tuning is a prevalent source of technical debt in LLM-based projects. For example, the comment \faCommenting{} \textcolor{gray}{\texttt{\# TODO: it might be profitable to fine-tune ... instead of letting the base model handle it}} suggests that developers acknowledge the potential performance improvements achievable through fine-tuning rather than relying solely on the base model. This observation also reflects a decreasing dependence on the prompt learning approach, as incorporating examples or context into a general-purpose model may yield the desired outcomes more effectively.

While prompt-learning appeared slightly less frequently in our dataset compared to fine-tuning, we observed often impact on Prompt Debt. For example, the comment \faCommenting{} \textcolor{gray}{ \texttt{\# TODO: consider few-shot examples}}. This case demonstrates a missing learning approach and restructuring the prompt design to include the example.

\begin{acmbox}
\noindent 
\textbf{Summary of RQ2:} Prompt design and configuration, hyperparameter tuning, and the integration of LLM frameworks are the primary sources of LLM-specific SATD.  
\end{acmbox}


\begin{table*}[b!]
\small
\centering
\caption{Distribution of Prompt Techniques, Associated SATD Examples, and Prompt Examples}
\label{tab:prompt_types}

\begin{tabular}{p{2.5cm} p{4.3cm} p{7.7cm} p{2cm}}
\toprule
\textbf{Technique} & \textbf{Prompt Example} & \textbf{Associated SATD} & \textbf{Count (\%)} \\
\midrule
Instruction 
& You create summaries that keep all the information from the original text. You must keep all numbers and statistics from the original text. You will provide the summary in succint bullet points. For longer inputs, summarise the text into more bullet points.....
& \# If the prompt is too long, warn and give up.\newline
\# TODO: Make it follow the prompt? Could be missing instructions\newline
\# I need the instructions very specific, but maybe I can break it up into multiple system prompts, and ......\newline
\#...... , you should revise the Instructions so that AI Assistant would quickly and correctly respond in the future.
& 66 (38.60\%) \\\hline

Few-Shot 
& Here are examples of different locations seperated by newlines ...
& \# TODO: make this few-shot on real examples instead of dummy ones\newline
\# TODO: run few-shot ... and positive/negative examples to determine final classification
& 31 (18.13\%) \\\hline

Doc 
& Summarize this document to answer this question: Document: \textit{PLACEHOLDER} If the document isn't relevant, answer: Not Relevant.
& \# todo: debug. "stuff" just adds all documents in the prompt. Test more smart approaches.\newline
\# TODO combine with LangChain.DocumentLoader?\newline
\# TODO: Optional chunking for documents that are too large
& 21 (12.28\%) \\\hline

Chain-of-Thought 
& ... Let's work this out in a step by step way to be sure we have the right report use the goal ...
& \# let's get a chain of thought (COT) approach to understanding the data set.\newline
\# TODO: after refactor: sample randomly instead, otherwise might e.g. only evaluate on CoT realized examples
& 12 (7.02\%) \\\hline

CodeBlock 
&  Review this diff code change and suggest possible improvements and issues, provide fix example... [\textit{PLACEHOLDER}]
& \# TODO: Consider adding project description for context in prompt\newline
\# TODO: make this less hardcoded with if-else statements
& 11 (6.43\%) \\\hline

Zero-Shot 
& What is a good title for this chat that is 20 characters or less?
& \# FIXME: Doesn't seem to have same max\_tokens == -1 for prompts==1
& 10 (5.85\%) \\
\bottomrule
\end{tabular}
\end{table*}

\vspace*{-5pt}
\subsection{\textbf{RQ3: Which prompt techniques are more prone to SATD?}}
For RQ3, we collected 155 developer-written prompts and categorized them based on the prompting techniques employed to identify which techniques are most prone to accumulating SATD. We analyzed the associated SATD comments for each technique to determine if they were directly related to the specific prompting approach used. Table \ref{tab:prompt_types} presents the results. In total, we analyzed 155 prompts, excluding 24 prompts (13.71\%) from classification due to a lack of meaningful content. For example, prompts like ['Hi, this is a test'] were categorized as unknown, as they did not provide sufficient context to be identified as any specific technique.

In examining the SATD comments associated with these prompt techniques, we found that a significant number of comments were specific to the techniques used. However, it is essential to note that not all SATD comments were directly related to the technique itself. For instance, SATD comments associated with Zero-shot prompts were more general and did not directly relate to the unique aspects of Zero-shot prompting. Our analysis highlights SATD comments directly referencing challenges inherent to the techniques.

{\bf Instruction Block} prompts, where developers provide explicit commands to LLMs, were identified as particularly vulnerable to SATD, comprising 37.7\% of our dataset. This technique often results in issues related to Prompt Debt, specifically suboptimal instruction clarity and excessive instruction length. Such issues can lead to comments like \faCommenting{}\textcolor{gray}{\textit{ \#If prompt is too long ...}} and \textcolor{gray}{\textit{.. Could be missing instructions}}, where developers acknowledge the need to adjust the prompt length and enhance the clarity of provided instructions. Instruction length is particularly problematic, as the model’s ability to retrieve and interpret instructions from the middle of the context diminishes as prompts become longer\cite{liu2024lost}. This can result in incomplete or less accurate outputs when essential details are buried within lengthy prompts. Unclear or suboptimal instructions, coupled with overly long prompt instructions, can severely impact the effectiveness of LLM responses by leading to misinterpretation or incomplete task execution. 

{\bf Few-shot} Few-shot prompts accounted for 17.7\% of our dataset, underscoring their common use in providing contextual examples to guide LLM behavior. The SATD comments linked to this technique were mainly associated with Learning Debt, as discussed in RQ2. These comments revealed key challenges, especially around improving the quality and relevance of the examples used. Comment like \faCommenting{} \textcolor{gray}{ \texttt{ make this few-shot on real examples ..}} suggests that developers often rely on placeholder examples that do not reflect real tasks. Similarly, \faCommenting{} \textcolor{gray}{ \texttt{ .. positive/negative examples to determine final classification}}. Without these carefully selected examples, classification task and model reliability may decline \cite{bhatia2023empirical,brown2020language}.

The {\bf Documentation} technique (12.28\%) involves incorporating documentation directly within prompts to provide context or background information for LLMs. SATD in this area often arises from the challenges of managing and processing large document inputs. Developers frequently encounter issues with the efficient handling of these substantial text blocks, as illustrated by comments like \faCommenting{} \textcolor{gray}{ \texttt{ .. adds all documents in the prompt. Test more smart approaches}} and \faCommenting{} \textcolor{gray}{ \texttt{Optional chunking for documents that are too large}}. These comments reflect the struggle to ensure that large documents are integrated in a way that maintains performance without overwhelming the model’s capacity.
To optimize prompt performance and reduce memory load, developers often identify the need for advanced document handling methods. For instance, comments like \faCommenting{} \textcolor{gray}{ \texttt{.. combine with LangChain.DocumentLoader?}} to integrating tools specifically designed for document management, which could improve the workflow.

The SATD examples for {\bf Chain-of-Thought (CoT)} and {\bf CodeBlock} prompt techniques were fewer than those for the aforementioned techniques, yet they reveal the challenges developers face when integrating these approaches. In CoT, issues often arise from a lack of refining and refactoring the step-by-step structure to improve response quality. In CodeBlock, challenges include providing insufficient code-related context within prompts and addressing hard-coded structures.

\begin{acmbox}
\noindent 
\textbf{Summary of RQ3:} Instruction Block prompts are the most prone to SATD, often due to prompt length and instruction clarity issues. Few-Shot prompts accumulate debt when placeholder examples are not replaced with relevant, task-specific ones. Documentation prompts face challenges with managing large text inputs.  
\end{acmbox}


\section{Additional Discussion and Implications} 
\label{sec:Study_Implication}
This section discusses the implications of our study's results in relation to each research question and provides practical guidance to mitigate technical debts arising from LLMs.

\vspace*{6pt}
\noindent
\textbf{Additional Discussions:}
In RQ1, we showed that LLM-based applications exhibit challenges that make developers prone to introducing technical debts. We found that OpenAI-based applications account for most of SATD in the LLM projects we investigated. This is likely attributable to both the widespread use of OpenAI's API by developers \cite{chen2024empirical} and the extensive number of associated comments (i.e., 197,380 out of 340,840 total comments). The combination of OpenAI with LangChain ranks second. This highlights that the interplay between LLM model functionalities and their orchestration is a critical factor in the accumulation of technical debt. This is also supported by our results in RQ2, LLM-Framework debts usually related to prompting design.

In RQ2, the most prevalent LLM-related debt was identified as Prompt Debt. Prompt engineering is an ongoing field that is continuously evolving both in academia and industry~\cite{fan2023large}. This rapid development creates challenges for developers, who are compelled to revisit and enhance their prompts as the design directly influences the desired output.
Our observations suggest that some of the Prompt Debt issues, such as length, not only influence the quality of the LLM output but also contribute to Cost Debt. Longer prompts contain more tokens, which, in turn, increase the cost of using LLM services. This is also similar to Hyperparameter Tuning Debt. We frequently encounter hyperparameters like max\_tokens, which set limits on the output length of LLMs. Although max\_tokens does not necessarily affect the diversity and creativity of the responses of the LLM, we believe that it could also influence \textit{Cost Debt}, as controlling the token limit affects the token pricing (i.e., a higher number of tokens increases the cost).

In RQ3, we found that Instruction Block, Few-shot, and Documentation prompts were the most prone to accumulating SATD, showing that certain prompting techniques naturally introduce specific risks. Instruction Block prompts often lead to issues with instruction clarity and length, Few-shot prompts with poor example selection, and Documentation prompts with large unstructured inputs that overwhelm context windows. However, some of this debt can be reduced by adopting alternative strategies; for instance, using Retrieval-Augmented Generation (RAG)~\cite{lewis2020retrieval} can help manage Documentation prompts more effectively by retrieving only relevant information instead of embedding entire documents.\\

\vspace*{6pt}
\noindent
\textbf{Implications:}

\vspace*{3pt}
\noindent
\textbf{Takeaway 1: Early Detection for Prompt Smells and Requirement Smells:}
Our findings on Prompt Debt closely relate to the ideas of Prompt Smells\cite{ronanki2024prompt} and Prompt Requirement Smells\cite{vogelsang2025impact}. Prompt Smells describe semantic and syntactic issues in prompts that reduce output desirability, explainability, or traceability, which can lead to lower output quality or harder-to-explain results. Prompt Requirement Smells, on the other hand, refer to issues like ambiguity or missing information when requirements are used as prompts, which can confuse the model and lower performance.

In our study, the issues we found under Prompt Debt—such as missing prompt templates, hardcoded values, long and unclear instructions, and poorly managed document inputs—align closely with Prompt Smells. These problems show that many prompts are fragile, difficult to maintain, and challenging for models to interpret reliably, directly affecting the quality of the outputs. 

At the same time, we also found cases that align more with Prompt Requirement Smells. For example, some comments showed developers struggling with selecting relevant few-shot examples or forgetting to include important project context in prompts (e.g., \faCommenting{} \texttt{\#TODO: Consider adding project description for context in prompt}). These issues mirror how vague or incomplete requirements can lead to requirement smells.

By identifying and addressing Prompt Debt early, developers can improve the clarity, structure, and completeness of prompts, which ultimately supports achieving the key aspects of generative AI output desirability.

\vspace*{3pt}
\noindent
\textbf{Takeaway 2: Towards Deterministic Hyperparameter Tuning:} The primary challenge lies in determining optimal threshold values for the hyperparameters of LLMs. Hyperparameters such as \texttt{temperature, top\_p, and top\_k} are critical because they significantly influence the diversity and creativity of the responses generated by the model~\cite{arora2024optimizing}. These settings adjust how the model predicts and varies its output. Developers should experiment with parameter values to identify the most suitable or deterministic thresholds for their specific tasks. Once these optimal thresholds are established, they can be consistently applied to similar tasks to maintain reliable performance. 

\noindent
\textbf{Takeaway 3: Managing LLM Framework Efficiently:} Frameworks like LangChain simplify prompt management and LLM output handling, but improper use can introduce technical debt. Our findings indicate that issues such as unstructured prompt construction and delayed adoption of output parsers contribute to inefficiencies. Developers should consistently use prompt templates to maintain structured and reusable prompts while integrating LangChain’s output parsers early to enforce predictable response formatting output.

\vspace*{3pt}
\noindent
\textbf{Takeaway 4: Cost-Aware LLM Selection:} Costs vary significantly among LLM providers. For instance, Wang et al.\cite{wang2024survey} show that using OpenAI’s GPT-4 can cost \$30 for 1 million tokens, whereas alternatives like Mistral 7B may cost as little as \$0.20. This difference emphasizes the importance for developers to carefully select models that align with their budget and performance requirements. Our findings indicate that \textit{Cost Debt} is a significant concern, with developers actively seeking ways to minimize costs. This highlights the importance of dynamically selecting from a range of LLMs, utilizing lower-cost for general tasks, and switching to more powerful, higher-cost models only when necessary.

\vspace*{3pt}
\noindent
\textbf{Takeaway 5: Improving Prompting Techniques:} 
Instruction Block and few-shot prompting techniques were the most prone to SATD. Excessive instruction length and poorly selected examples can reduce prompt clarity and output quality. Developers should focus on concise, well-structured instructions and use example selection methods (e.g., embedding-based similarity~\cite{steenhoek2024comprehensive}) to select more relevant few-shot examples, thereby improving model performance.

\section{Threats To Validity}
\label{sec:ThreatsToValidity}
\vspace*{3pt}
\textbf{Internal Validity:} 
One potential threat to our internal validity lies in the accuracy of the tool (SATDDetector~\cite{obrien202223}) used to detect SATD in LLM-focused repositories. Although SATDDetector has been effective in prior studies, it was originally designed for more general software settings. To mitigate this risk, we supplemented the tool’s results with a keyword-based detection approach\cite{da2017using}. Another concern involves human bias in the manual classification of LLM-specific debts (RQ2). By following a process well established in prior work\cite{obrien202223}, we used a coding process for classification, achieving a Cohen’s Kappa of 0.74 (indicating substantial agreement). For our GPT-based prompt classification, we additionally validated a random sample of 40 instances (26\% of the dataset), finding a 92.5\% correct classification rate. This approach reduces subjective errors and increases confidence in our classification outcomes. We chose not to use an LLM-based classifier to identify LLM-related debt because this domain is still emerging and frequently requires human judgment. For instance, automated methods might fail to recognize certain framework-specific debts (Section 3.4) if the classification categories are not yet well established. In contrast, manual classification allows for more precise detection of these emerging debt types.

\vspace*{3pt}
\textbf{External Validity:} Our findings may be limited in scope because they focus on LLMs based primarily on PromptSet projects\cite{pister2024promptset}, which capture a specific subset of LLM-based projects and LLM frameworks. Consequently, the types of SATD identified might not apply equally to highly customized transformer models, where deeper configurations—such as learning rates, optimizer choices, batch sizes, and fine-tuning epochs—are more prominent. In contrast, general-purpose LLMs emphasize prompt engineering and API-level customization. Similarly, our findings on LLM-Integrated Framework Debt primarily focus on LangChain due to its widespread use. Still, similar challenges may arise in other LLM orchestration frameworks like LlamaIndex, which also introduce complexities in prompt management, response parsing, and dependency maintenance. Future studies should expand to a wider range of models and LLM frameworks to ensure that the forms of SATD uncovered generalize across diverse LLM-based applications.

\vspace*{3pt}

\textbf{Construct validity:}  
This concerns the extent to which our study accurately captures LLM-related SATD. We refined our classification by integrating insights from official LLM documentation~\cite{openai2023, langchain2023, cohere2023, anthropic2023} and empirical research on LLM-related development challenges~\cite{chen2024empirical, hou2024voices, tafreshipour2024prompting}. Additionally, we considered inference-related debt~\cite{pepe2024taxonomy}, which briefly addresses prompt formulation and hyperparameter tuning within a deep learning (DL) applications.
Moreover, to improve construct clarity, we distinguished between closely related debt types.
For example, while few-shot examples are embedded in prompts and often overlaps with prompt formulation, we categorized comments as Learning Debt when they involved reasoning about example selection or task-specific adaptation. This distinction allowed us to separate issues related to high-level learning approach from those related to Prompt Debt.

\section{Conclusion And Future Work}
\label{sec:Conclusion}
This paper presents the first empirical investigation into self-admitted technical debt (SATD) specific to large language models (LLMs). Analyzing 93,142 Python files, the study finds that OpenAI accounts for 54.49\% and the LangChain framework 12.35\% of SATD instances. Prompt configuration is identified as the primary source of LLM-specific SATD, with instruction-based (37.7\%) and few-shot prompts (17.7\%) being the most prone to prompting techniques due to the lack of instruction clarity and example quality. The study provides a reproducible dataset and offers insights on LLM-related technical debt for more maintainable applications. In the future, we plan to expand our exploration of technical debt associated with LLMs by encompassing a wider range of debt types. Additionally, we aim to investigate the specific technical debts linked to various prompt techniques.

\bibliographystyle{ACM-Reference-Format}
\bibliography{Bib/bib}

\newpage
\centering
{\bf Appendix A}

\vspace{10pt}


\begin{minipage}{0.95\linewidth} 
\begin{lstlisting}[label={lst:llm_keywords}]
# Keywords for Identifying LLMs
OpenAI: [
    "openai", 
    "OpenAI", 
    "openai.Completion.create", 
    "openai.ChatCompletion.create", 
    "openai.Completion", 
    "openai.ChatCompletion"],
Anthropic: [
    "anthropic", 
    "Claude", 
    "anthropic.Completion.create", 
    "claude.Completion"],
Cohere: [
    "cohere", 
    "cohere.Client", 
    "cohere.generate", 
    "cohere.chat", 
    "cohere.summarize"],
LangChain: [
    "langchain", 
    "LLMChain", 
    "PromptTemplate", 
    "HumanMessage", 
    "AIMessage", 
    "BaseTool", 
    "@tool", 
    "langchain.llms"]
\end{lstlisting}
\end{minipage}
\end{document}